\documentclass[table,xcdraw,dvipsnames,svgnames]{article}
\usepackage{aaai2026}  
\usepackage{times}  
\usepackage{helvet}  
\usepackage{courier}  
\usepackage[hyphens]{url}  
\usepackage{graphicx} 
\usepackage{natbib}  
\usepackage{caption} 
\usepackage{algorithm}
\nocopyright

\def\frameworkname{SoundCLIP}   
  
\usepackage{booktabs}
\usepackage{array}
\usepackage{multirow}
\usepackage{graphicx}
\usepackage{colortbl}

\definecolor{headerblue}{RGB}{102, 126, 234}
\definecolor{lightblue}{RGB}{240, 248, 255}
\definecolor{improvementblue}{RGB}{52, 152, 219}
\definecolor{lightimprovement}{RGB}{235, 245, 255}
\definecolor{darkgray}{RGB}{52, 73, 94}
\definecolor{rawcolor}{RGB}{231, 76, 60}
\definecolor{projectedcolor}{RGB}{39, 174, 96}
\definecolor{headerblue}{RGB}{102, 126, 234}
\definecolor{lightblue}{RGB}{240, 248, 255}
\definecolor{rawgreen}{RGB}{39, 174, 96}
\definecolor{lightrawgreen}{RGB}{235, 250, 235}
\definecolor{projectedred}{RGB}{231, 76, 60}
\definecolor{lightprojectedred}{RGB}{255, 235, 235}
\definecolor{darkgray}{RGB}{52, 73, 94}
\definecolor{metricheader}{RGB}{149, 165, 166}
\definecolor{headerblue}{RGB}{102, 126, 234}
\definecolor{lightblue}{RGB}{240, 248, 255}
\definecolor{darkgray}{RGB}{52, 73, 94}
\usepackage{tcolorbox}
\usepackage{pifont}
\usepackage{amsthm}
\usepackage{amsthm}

\definecolor{lightorange}{RGB}{255,200,100} 
\usepackage{colortbl}
\definecolor{lightgreen}{RGB}{230,245,230}
\definecolor{lightred}{RGB}{255,230,230}
\definecolor{lightblue}{RGB}{220,235,250}
\definecolor{lightgray}{RGB}{240,240,240}
\definecolor{lavender}{RGB}{230,230,250}
\definecolor{wheat}{RGB}{245,222,179}
\definecolor{neuripsblue}{rgb}{0.21,0.49,0.74}  
\definecolor{ftgray}{RGB}{245,248,250}          

\usepackage{graphicx}    
\usepackage{svg}         
\usepackage{pgf}
\usepackage{pgfplots}
\usepackage{tikz}
\usetikzlibrary{
  positioning,
  arrows,
  arrows.meta,
  calc,
  shapes.geometric,
  fit,
  backgrounds,
  decorations,
  decorations.pathmorphing
}

\usepackage{amsmath}     
\usepackage{amssymb}     
\usepackage{mathtools}   
\usepackage{textgreek}

\usepackage{algpseudocode}

\usepackage{multirow}    
\usepackage{booktabs}    
\usepackage{array}
\usepackage{threeparttable}
\usepackage{siunitx}     
\usepackage{adjustbox}   
\usepackage{subcaption}  
\usepackage{caption}     

\usepackage[utf8]{inputenc} 
\usepackage[T1]{fontenc}    

\usepackage{url}            
\usepackage{booktabs}       
\usepackage{amsfonts}       
\usepackage{nicefrac}       
\usepackage{svg}            
\usepackage{microtype}      
\usepackage{enumitem}
\usepackage{amsmath}

\usepackage{caption}
\usepackage{graphicx}
\usepackage{arydshln} 
\usepackage{booktabs}
\usepackage{multirow}
\usepackage{adjustbox}

\usepackage{newfloat}
\usepackage{listings}
\DeclareCaptionStyle{ruled}{labelfont=normalfont,labelsep=colon,strut=off} 
\floatstyle{ruled}
\newfloat{listing}{tb}{lst}{}
\floatname{listing}{Listing}
\title{\raisebox{-1.0ex}{\includegraphics[height=3.5ex]{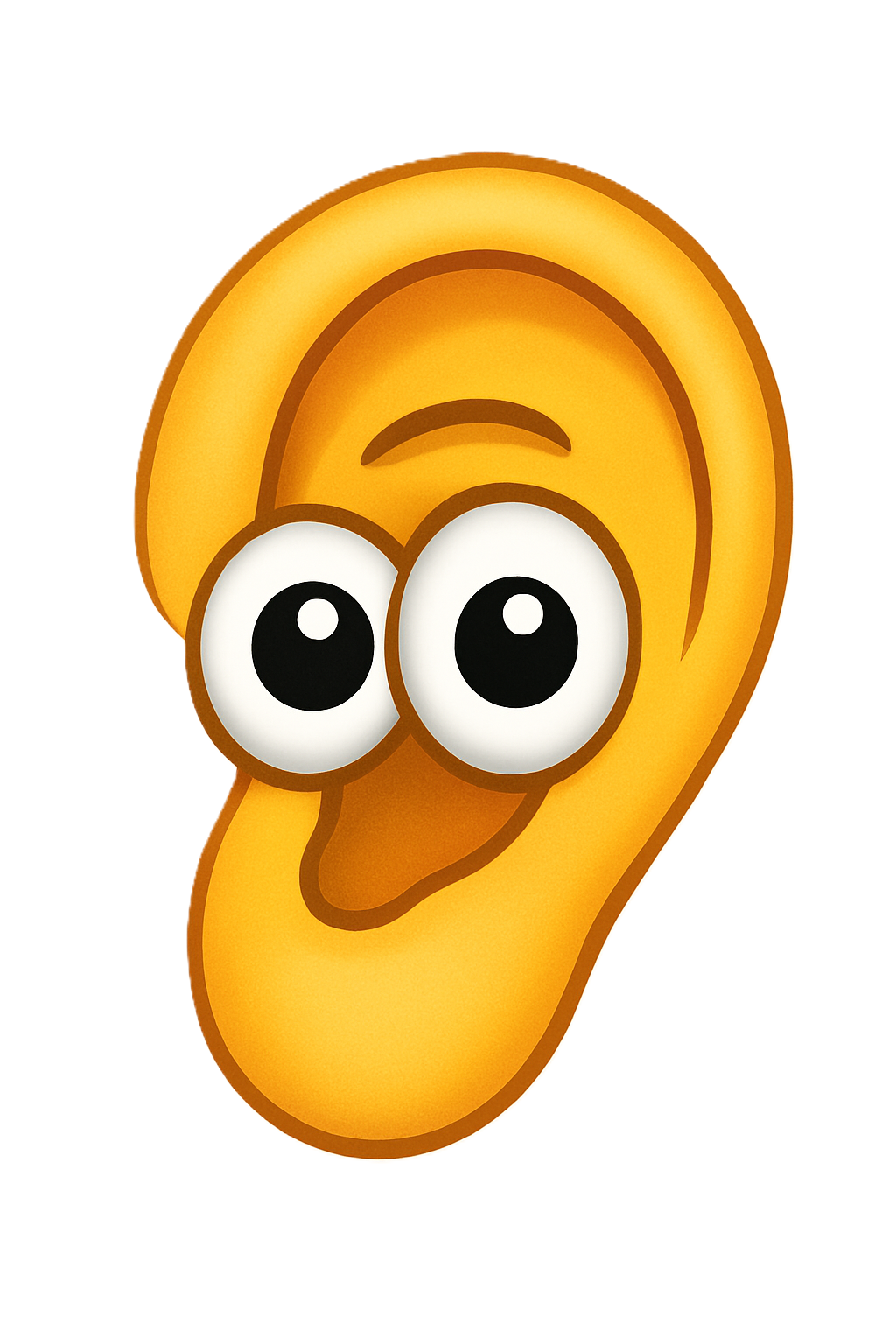}}\hspace{0.5em}Can Sound Replace Vision in LLaVA With Token Substitution?}

\author{
    Ali Vosoughi,
    Jing Bi,
    Pinxin Liu,
    Yunlong Tang,
    Chenliang Xu
}
\affiliations{
    Computer Science Department, University of Rochester, 
    NY, USA\\
    \{ali.vosoughi, jing.bi, pliu23, yunlong.tang, chenliang.xu\}@rochester.edu\\
    \textbf{Project Website:} https://ali-vosoughi.github.io/SoundCLIP/
}

\usepackage{bibentry}

\begin{document}

\maketitle

\begin{abstract}

What happens when we push audio-visual alignment to its absolute limits? To systematically investigate this question, we needed datasets with granular alignment quality annotations, but existing datasets treat alignment as binary, either synchronized or not. To address this limitation, we developed a comprehensive dataset featuring detailed alignment scores that reveal the hidden spectrum of audio-visual perceptual correspondence. Using these precise scores, we create "superaligned" representations by training exclusively on the most perfectly matched audio-visual pairs, then conduct our systematic investigation into how this extreme alignment transforms perceptual model behavior across retrieval and generation tasks.
The encoders under study fall into two main groups consisting of image-centric encoders that were pretrained using visual modalities as intermediary hubs for connecting modalities, and text-centric encoders that were pretrained with direct audio-language alignment.
We first measure the baseline performance of these encoders on two key tasks, namely cross-modal retrieval and text description generation in vision-language models. Subsequently, we realign all encoders with the CLIP space using highly coherent audio-visual data and observe the performance changes.
Our findings reveal that the initial architectural type of the encoder determines how it responds to the alignment process. Image-centric encoders, which are inherently designed for alignment, demonstrate exceptional performance in cross-modal retrieval, but this intensive alignment causes compression of unique linguistic information and reduces the quality of their text description generation in vision-language models.
In contrast, text-centric encoders, which possess stronger linguistic authenticity, are able to maintain a better balance between the two objectives. This pattern reveals a fundamental trade-off where excessive alignment with the visual manifold leads to improved retrieval capabilities, but simultaneously reduces the richness of acoustic and linguistic information necessary for quality text description generation.
Image-centric encoders are preferable for retrieval tasks, while text-centric encoders are preferable for text description generation in vision-language systems. The results also demonstrate that in designing multimodal systems, one must consider the inherent trade-off between alignment depth and preservation of domain-specific information.

\end{abstract}

\section{Introduction}

Concurrent with psychophysical discoveries of multisensory integration, early computational models such as Hidden Markov Models (HMMs) \citep{bengio2002ahmm} sought an algorithmic fusion of auditory and visual streams. With the recent rise of large-scale multimodal pre-training, it has become possible to learn representations that connect vision, language, and sound. CLIP demonstrates that contrastive learning on 400 million image–text pairs produces general-purpose visual features \citep{radford2021learning}. Inspired by its success, researchers have also extended CLIP to include audio, training audio encoders alongside vision language objectives. AudioCLIP adds an audio branch to CLIP that is trained on AudioSet and enables unsupervised sound classification and cross-modal retrieval \citep{guzhov2022audioclip}. Wav2CLIP distils CLIP into a purely audio encoder by predicting sound in CLIP's embedding space \citep{wu2022wav2clip}, and ImageBind learns a single embedding space that uses images as a hub to tie audio to five other modalities \citep{girdhar2023imagebind}. CLAP takes a different approach and aligns audio with language through contrastive audio–text pre-training \citep{elizalde2023clap}, whereas Whisper learns rich audio representations from large-scale supervised training \citep{radford2023whisper}. 

\begin{figure}[!t]
    \centering
    \includegraphics[width=0.6\linewidth]{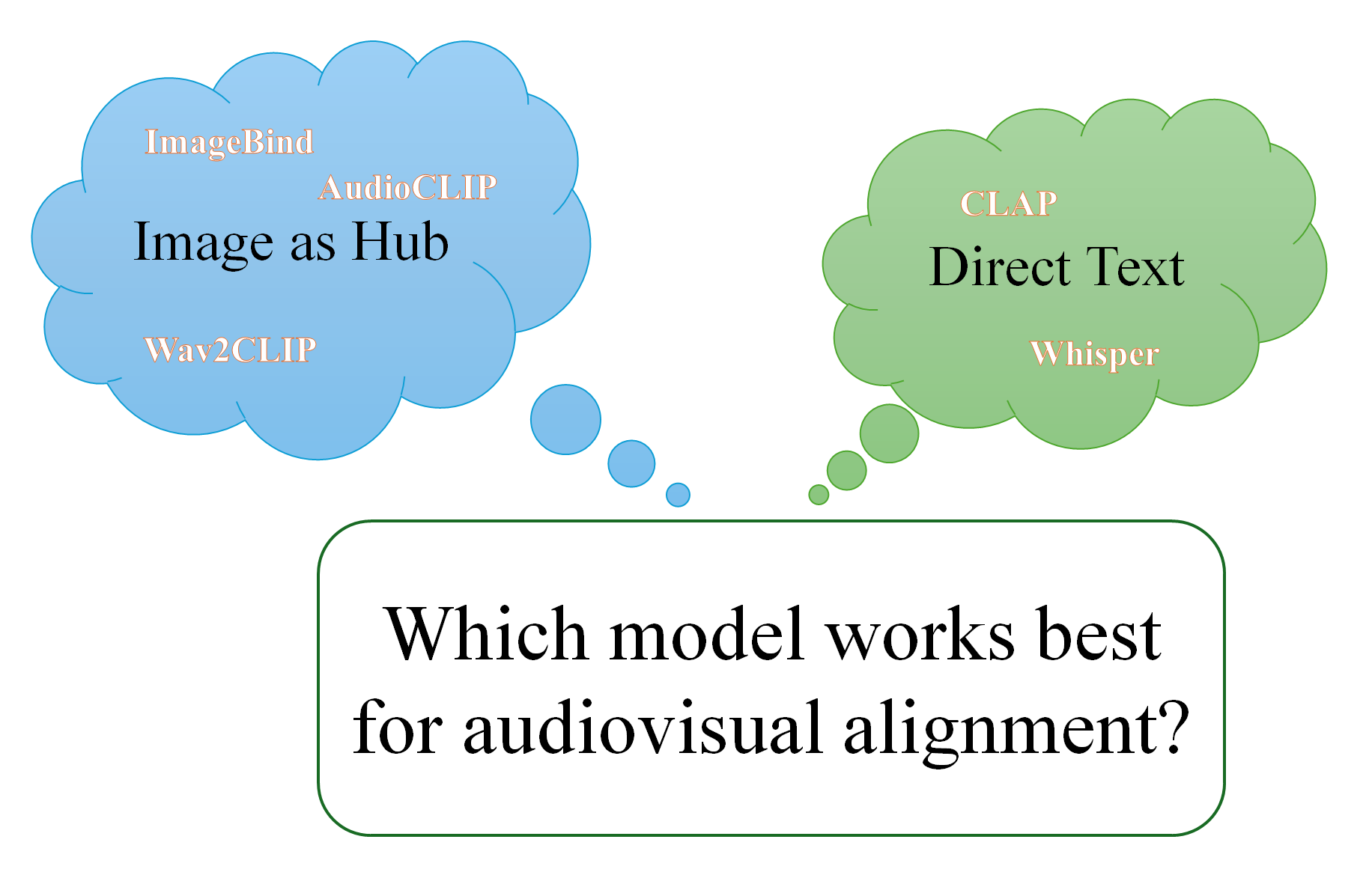}
    \caption{Audio encoders fall into two distinct camps: image-centric models (left) that use images as central hubs for connecting modalities, and text-centric models (right) that directly align audio and language. When the goal is audio--visual alignment, which approach proves superior?}
    \label{fig:teaser}
\end{figure}

From the perspective of paired training data, audio encoders fall into two distinct camps: \emph{image-centric models}, which use pictures as central hubs for modality binding, and \emph{text-centric models}, which align audio directly with language. Unfortunately, none of these works has examined which family of encoders performs better when audio-visual alignment is required in semantic space. In this work, we explicitly exploit this duality. The clustering of the five audio encoders we study reveals image-centric models that link modalities through vision and text-centric models that align audio directly with language, as shown in Fig. \ref{fig:teaser}. This conceptual divide motivates our investigation into how forcing audio embeddings toward a fixed visual manifold affects retrieval and generation tasks. Concretely, we freeze the visual CLIP encoder and study how different audio encoders behave when their embeddings are projected into the CLIP space. We evaluated five representative encoders that cover the two training paradigms. Language-supervised models such as CLAP and Whisper are trained in audio–text pairs and embed audio in the semantic space \citep{elizalde2023clap,radford2023whisper}. Vision-centric encoders—including AudioCLIP, Wav2CLIP, and ImageBind—learn audio features through alignment with images; AudioCLIP is jointly trained with images and text to extend CLIP \citep{guzhov2022audioclip}, Wav2CLIP distils audio from CLIP's visual branch \citep{wu2022wav2clip}, and ImageBind uses images as the natural hub for cross-modal binding \citep{girdhar2023imagebind}. By comparing these paradigms, we can disentangle the effects of language supervision vs. vision-based alignment on downstream tasks.

To precisely evaluate audio–visual alignment, we introduce \textbf{AVE-2}, a dataset of 580,145 clips with five-dimensional annotations that measure temporal accuracy, spatial coherence, contextual relatedness, physical causality, and source visibility. Unlike the original AVE dataset, which treats synchrony as a binary attribute \citep{tian2018ave}, AVE-2 quantifies how well sound and video match along multiple axes and provides a test-bed for studying continuous alignment and its impact on retrieval and generation. We also develop a lightweight fusion framework that injects audio tokens without architectural changes into LLaVA \citep{liu2023visual}. Two strategies are explored: (i) predicting the audio embedding into CLIP space using a small multilayer perceptron, and (ii) padding raw audio features to match the visual token dimension. Our experiments reveal a previously unknown trade-off between retrieval and generation. Aligning audio to the CLIP visual manifold dramatically improves retrieval but at the cost of reduced caption quality; raw embeddings produce richer descriptions, but yield weaker retrieval performance. Moreover, we observe that language-supervised encoders (Whisper and CLAP) retain better generative capabilities than vision-centric encoders even under forced alignment. To this end, we also propose \textbf{WhisperCLIP}, which combines intermediate Whisper layers' features and offers a favourable balance between alignment and generation quality. 

The remaining sections present the dataset and statistics, the design of the \emph{SoundCLIP} framework and the proposed \emph{WhisperCLIP} method in the \emph{Methodology} section, followed by extensive experiments for detailed evaluations. Related work is discussed before the conclusion.
\section{AudioVisual Event Evaluation (AVE-2) Dataset}
\label{sec:dataset}

We began with a fundamental question: what happens when audio and visual information are perfectly synchronized? Does it truly make a difference for artificial intelligence models? To answer this, we needed data where we could precisely measure how well-aligned each audio-visual pair was. However, when we examined existing datasets, we discovered a critical limitation. Every data set treated alignment as a simple binary question—either synchronized or not synchronized. It was like asking whether the weather is hot or cold without actually measuring the temperature.

\begin{table*}[!t]
\centering
\scriptsize
\definecolor{royalblue}{RGB}{65,105,225}
\definecolor{academicgray}{RGB}{245,245,245}
\begin{tabular}{>{\columncolor{academicgray}}p{3.2cm}*{6}{p{1.8cm}}}
\toprule
\rowcolor{royalblue}
\textcolor{white}{\textbf{Capability}} & \textcolor{white}{\textbf{AVE-2}} & \textcolor{white}{\textbf{AudioSet}} & \textcolor{white}{\textbf{VGG-Sound}} & \textcolor{white}{\textbf{AVE}} & \textcolor{white}{\textbf{Music-MIT}} & \textcolor{white}{\textbf{EPIC-Kit.}} \\
\midrule
\textbf{Alignment Study} & \textcolor{royalblue}{\ding{51}} \textbf{5D scores} & \textcolor{red}{\ding{55}} None & \textcolor{red}{\ding{55}} None & \textcolor{orange}{\ding{115}} Binary & \textcolor{orange}{\ding{115}} Limited & \textcolor{red}{\ding{55}} None \\
\rowcolor{academicgray}
\textbf{Temporal Precision} & \textcolor{royalblue}{\ding{51}} \textbf{±40ms} & \textcolor{red}{\ding{55}} ~1s & \textcolor{red}{\ding{55}} ~1s & \textcolor{orange}{\ding{115}} ~1s & \textcolor{royalblue}{\ding{51}} Frame & \textcolor{red}{\ding{55}} ~1s \\
\textbf{Invisible Sources} & \textcolor{royalblue}{\ding{51}} \textbf{337K} & \textcolor{orange}{\ding{115}} Few & \textcolor{orange}{\ding{115}} Few & \textcolor{red}{\ding{55}} None & \textcolor{red}{\ding{55}} None & \textcolor{red}{\ding{55}} None \\
\rowcolor{academicgray}
\textbf{Causality Analysis} & \textcolor{royalblue}{\ding{51}} \textbf{Yes} & \textcolor{red}{\ding{55}} No & \textcolor{red}{\ding{55}} No & \textcolor{red}{\ding{55}} No & \textcolor{red}{\ding{55}} No & \textcolor{orange}{\ding{115}} Partial \\
\textbf{Quality Control} & \textcolor{royalblue}{\ding{51}} \textbf{Multi-level} & \textcolor{red}{\ding{55}} None & \textcolor{red}{\ding{55}} None & \textcolor{red}{\ding{55}} None & \textcolor{red}{\ding{55}} None & \textcolor{red}{\ding{55}} None \\
\rowcolor{academicgray}
\textbf{Score Type} & \textcolor{royalblue}{\ding{51}} \textbf{Continuous} & \textcolor{red}{\ding{55}} Category & \textcolor{red}{\ding{55}} Category & \textcolor{orange}{\ding{115}} Binary & \textcolor{orange}{\ding{115}} Limited & \textcolor{red}{\ding{55}} Category \\
\bottomrule
\end{tabular}
\caption{Comparison with existing datasets showing AVE-2's unique capabilities for alignment research.}
\label{tab:dataset_comparison}
\end{table*}

How could we study whether perfect temporal synchronization matters more than spatial coherence? How could we investigate whether physical causality plays a role in understanding complex scenes? The existing datasets simply couldn't provide these answers because they lacked the granular measurements we needed for systematic investigation. So we decided to create our own solution. We started systematically evaluating every three-second video clip using three advanced models with specialized expertise in multimodal signal analysis and perceptual psychology. Instead of simply saying "aligned" or "not aligned," we leveraged a five-dimensional scoring system that gave each video segment a detailed quality profile as described in  [https://avva-curation.github.io/AVVA-curation/] \citep{vosoughi2025quality}, allowing us to understand exactly where each clip excelled or struggled (Table~\ref{tab:alignment_dimensions}).

\begin{table}[htbp]
\centering
\scalebox{0.64}{
\begin{tabular}{lcc}
\toprule
\textbf{Alignment Dimension} & \textbf{Mean Score} & \textbf{Model Impact} \\
\midrule
Temporal Alignment & 7.54 & Event detection \& sequence prediction \\
Spatial Coherence & 8.35 & Source localization \& spatial understanding \\
Contextual Relevance & 6.32 & Scene comprehension \& context inference \\
Physical Causality & 9.02 & Cause-effect relationship recognition \\
Source Visibility & 8.04 & Out-of-frame reasoning capabilities \\
\bottomrule
\end{tabular}
}
\caption{Our five-dimensional scoring system measuring different aspects of audio-visual alignment quality.}
\label{tab:alignment_dimensions}
\end{table}

When this rigorous annotation process was complete, we had created something unprecedented: 580,147 video segments, each with its own comprehensive five-dimensional quality assessment. The scoring revealed fascinating patterns that had never been observed before. Physical Causality demonstrated remarkable consistency with minimal variance (standard deviation = 0.51, median = 9.0), while Contextual Relevance showed much greater variability (standard deviation = 1.33, median = 6.0), reflecting the inherent complexity of semantic assessment in real-world scenarios.

Our approach differs fundamentally from existing datasets, as demonstrated by systematic comparison (Table~\ref{tab:dataset_comparison}). Although AudioSet and VGG-Sound \citep{chen2020vggsound} provide impressive volume, they do not offer alignment measurements. The AVE dataset \citep{tian2018ave} provides only binary presence labels. Music-MIT offers frame-level precision, but focuses exclusively on musical content. EPIC-Kitchens \citep{damen2020epic} emphasizes action recognition without considering alignment quality. Only AVE-2 enables researchers to systematically investigate how alignment quality affects model performance.

The final dataset achieves strategic balance across three critical content types that enable comprehensive analysis. We have 490,546 segments with visible sound sources, perfect for studying direct audio-visual alignment, averaging 1.08 visible sources per segment. Another 337,069 segments contain invisible sound sources that challenge models to reason beyond what they can see, averaging 0.63 invisible sources per segment and testing out-of-frame reasoning capabilities. Additionally, 446,567 segments show silent objects, averaging 1.00 silent objects per segment and helping us understand when models incorrectly expect sounds that aren't there.

This comprehensive coverage spans from simple single-source scenarios—62,073 segments containing only visible sources (10.7\% of the dataset)—to complex multi-source environments where all three content types coexist simultaneously in 227,677 segments (39.2\% of the dataset). The remarkable diversity enables researchers to conduct controlled experiments that were previously impossible with existing datasets.

Researchers can now train models exclusively on clips with perfect temporal alignment but moderate spatial coherence, measuring how each dimension independently contributes to performance. They can identify optimal combinations of alignment qualities for different tasks, systematically reveal model weaknesses in processing invisible sources or complex causal relationships, and perform regression analysis using our continuous scoring system rather than being limited to categorical classifications.

The comprehensive coverage enables statistical analyses impossible with existing datasets. Content is distributed across 157,800 unique visible source categories, 111,644 invisible source categories, and 300,265 silent object categories, providing sufficient diversity for meaningful statistical analysis and linear correlation studies. The continuous nature of our five-dimensional scoring system supports sophisticated analytical approaches that include regression analysis, correlation studies, and graduated quality thresholds.

\section{Methodology}
\label{sec:method}

\begin{figure*}[!t]
    \centering
    \includegraphics[width=0.9\linewidth]{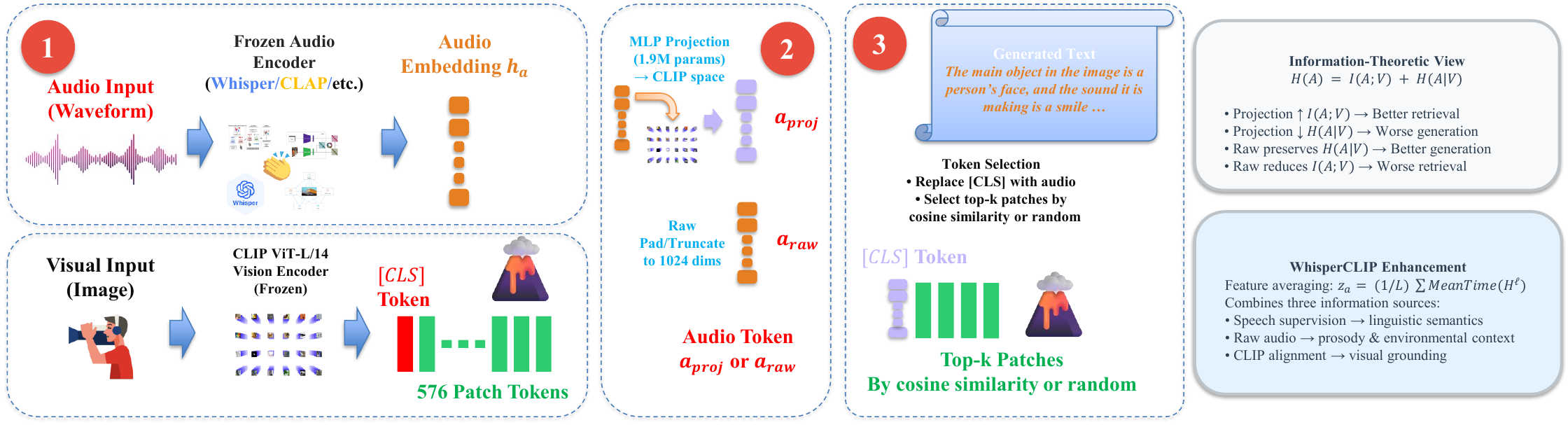}
    \caption{\textbf{SoundCLIP Architecture.} The pipeline consists of: (1) encoding audio with a frozen encoder, (2) optionally projecting this embedding into CLIP's visual space with a lightweight MLP (1.9M parameters) or preserving the raw embedding, and (3) replacing the visual \texttt{[CLS]} token with the audio token while selecting  $k$ visual tokens according to cosine similarity.}
    \label{fig:framework_overview}
\end{figure*}

A long–term goal for multimodal understanding is to let language models reason about scenes using sound alone.  In principle, we would like to discard all vision tokens and replace them with a single audio token that conveys everything the model needs to know about the scene.  Achieving this would require \emph{post‑training} the language model on additional instruction‑following data so that it learns how to interpret purely acoustic context.  Such retraining is beyond the scope of this work.  Instead, our study asks a more targeted question: how do different audio encoders behave when forced to align as tightly as possible with visual features versus when they remain independent?  

We consider the CLIP ViT–L/14 encoder, whose output consists of a global \texttt{[CLS]} token and 576 patch tokens.  In the visual domain, contrastive pre‑training encourages alignment between the \texttt{[CLS]} token and the patch tokens, so that the \texttt{[CLS]} summarises the content of the image.  By analogy, we would like to use a single audio token in place of the \texttt{[CLS]} token.  Formally, let $\boldsymbol{w}=(w_1,\dots,w_L)$ denote a text sequence and let $a$ be the audio token.  In an autoregressive language model, the conditional probability of the text given only this audio context is
\begin{equation}
P(\boldsymbol{w}\mid a)\;=\;\prod_{t=1}^L P\bigl(w_t\mid a,w_{<t}\bigr).
\label{eq:clsonly}
\end{equation}
In practice, due to the projection layers that connect visual embeddings to the language model’s token space, the model collapses if all patch tokens are removed.  To find the practical limits, we determine empirically a minimum number of patch tokens, denoted $k_{\mathrm{crit}}$, below which the language model outputs degenerate, random‑looking strings.  At this edge of collapse, adding any additional context is beneficial.
Accordingly, in our experiments we condition the model on the audio token $a$ and a small set of $k$ visual tokens $\{v_{i_1},\dots,v_{i_k}\}$:
\begin{equation}
P(\boldsymbol{w}\mid a, v_{i_1},\dots,v_{i_k})\;=\;\prod_{t=1}^L P\bigl(w_t\mid \{a,v_{i_1},\dots,v_{i_k}\},w_{<t}\bigr).
\label{eq:minpatch}
\end{equation}
We explore two natural strategies for choosing these $k$ tokens: selecting them uniformly at random, or selecting the $k$ patches whose embeddings are most similar to the audio token according to cosine similarity.  The latter focuses the model’s attention on visual regions most related to the sound.

\subsection{Information–Theoretic Perspective}

The alignment between the audio \texttt{CLS} token and the CLIP \texttt{CLS} token encourages the model to learn shared semantic features.  Let $A$ denote the random variable associated with the audio embedding and $V$ the variable associated with the visual embedding.  The total information in $A$ decomposes into two parts via the chain rule for entropy:
\begin{equation}
H(A)\;=\;I(A;V)\; +\;H(A\mid V).
\end{equation}
Here $I(A;V)$ measures the information common to both modalities, while $H(A\mid V)$ measures the modality‑specific information in audio that cannot be inferred from vision.  Any deterministic projection $f(A)$ used to align audio to the visual space cannot increase $I(A;V)$ and will inevitably decrease $H(A\mid V)$.  In other words, increasing cross‑modal alignment improves the model’s ability to find correspondences (retrieval) but reduces the richness of modality‑specific details needed for generation.  This trade‑off is central to our design: models trained to maximise $I(A;V)$ excel at retrieval, whereas models that preserve $H(A\mid V)$ yield higher‑quality captions.

\subsection{Alignment Strategies}

Our framework considers two ways of constructing the audio token $a$ from the encoder output $\mathbf{h}_a$.

\paragraph{CLIP‑aligned (projection).}  We train a small multilayer perceptron $\mathcal{M}$ to map $\mathbf{h}_a$ into the CLIP visual embedding space.  The resulting token $a_{\mathrm{proj}}=\mathcal{M}(\mathbf{h}_a)$ aligns closely with visual embeddings, maximising $I(A;V)$ and enabling the model to retrieve matching images.  However, this projection discards modality‑specific acoustic nuances, which we quantify using lower caption quality.
\paragraph{Raw (padding).}  To preserve the native geometry of the audio embedding, we simply pad or truncate $\mathbf{h}_a$ to 1024 dimensions.  The resulting token $a_{\mathrm{raw}}$ retains more of $H(A\mid V)$ but is less aligned with the visual space.  This approach favours generation tasks at the expense of retrieval performance.  Notably, audio encoders trained with natural‑language supervision, such as Whisper, CLAP and ImageBind, already embed linguistic meaning in their representations.  When used with the raw strategy, these encoders strike a better balance between shared and unique information than purely vision‑aligned encoders.

\subsection{WhisperCLIP and Multimodal Synergy}

To further improve the richness of the audio token, we propose \emph{WhisperCLIP}.  Whisper encoders learn rich acoustic representations from speech; however, deeper layers of Whisper emphasize transcription and gradually discard non‑speech information.  We compute an averaged representation across all layers to preserve both speech and environmental sounds:
\begin{equation}
\mathbf{z}_a\;=\;\frac{1}{L}\sum_{\ell=1}^{L}\operatorname{MeanTime}\bigl(\mathbf{H}^{(\ell)}\bigr),
\end{equation}
where $\mathbf{H}^{(\ell)}$ are the hidden states and $\operatorname{MeanTime}$ averages over time.  This fused representation is then used either as $a_{\mathrm{raw}}$ or projected via $\mathcal{M}$ to obtain $a_{\mathrm{proj}}$.

WhisperCLIP benefits from three sources of information.  Speech supervision endows its embeddings with linguistic semantics; the raw audio signal carries prosody and environmental context; and alignment with CLIP introduces visual grounding.  By combining these, WhisperCLIP offers a synergistic audio token that can support retrieval, captioning and question‑answering without additional post‑training.

\subsection{Practical Integration and Token Selection}

The final component of SoundCLIP is token substitution.  Given the audio token $a$ and patch embeddings $\{\mathbf{v}_1,\dots,\mathbf{v}_{576}\}$, we replace the visual \texttt{[CLS]} token with $a$ and select $k$ patches to retain.  Equation\,(\ref{eq:minpatch}) defines the autoregressive likelihood under this mixed context.  We choose the $k$ patches either at random or by selecting those with highest cosine similarity to $a$.  The former imposes no bias; the latter explicitly focuses on regions most related to the sound.  The optimal $k$ depends on the task: smaller $k$ relies more on audio for generation, whereas larger $k$ is needed for more visual context and visual tokens for generation.

\subsection{Implications for Language Models}

The information–theoretic analysis predicts—and our simulations confirm—that maximising $I(A;V)$ via projection improves retrieval but harms generation, while preserving $H(A\mid V)$ via padding improves generation at the cost of retrieval.  Equations\,(\ref{eq:clsonly}) and~(\ref{eq:minpatch}) illustrate how conditioning the language model on different contexts changes the distribution over outputs.  When conditioned only on the audio token (Eq.\,\ref{eq:clsonly}), the model fails because it lacks grounding; when conditioned on the audio token plus a minimal set of visual tokens (Eq.\,\ref{eq:minpatch}), it can maintain coherence.  These findings support the broader contribution of SoundCLIP: by carefully choosing the form of the audio token and the number of visual patches, one can traverse the trade‑off between cross‑modal retrieval and text generation and exploit the synergy of semantics, audio and vision.

\section{Experiments}
\label{sec:experiments}

\begin{table*}[!t]
\centering
\scalebox{0.46}{
\setlength{\tabcolsep}{0.9pt}
\setlength{\arrayrulewidth}{0.8pt}
\renewcommand{\arraystretch}{1.3}
\begin{tabular}{>{\columncolor{headerblue}}l|>{\columncolor{lightblue}}c>{\columncolor{lightblue}}c>{\columncolor{lightblue}}c|>{\columncolor{lightblue}}c>{\columncolor{lightblue}}c>{\columncolor{lightblue}}c|>{\columncolor{lightblue}}c>{\columncolor{lightblue}}c>{\columncolor{lightblue}}c|>{\columncolor{lightblue}}c>{\columncolor{lightblue}}c>{\columncolor{lightblue}}c|>{\columncolor{lightblue}}c>{\columncolor{lightblue}}c>{\columncolor{lightblue}}c|>{\columncolor{lightblue}}c>{\columncolor{lightblue}}c>{\columncolor{lightblue}}c|>{\columncolor{lightblue}}c>{\columncolor{lightblue}}c>{\columncolor{lightblue}}c|>{\columncolor{lightblue}}c>{\columncolor{lightblue}}c>{\columncolor{lightblue}}c}
\toprule
\multicolumn{1}{>{\columncolor{headerblue}}c|}{\textcolor{white}{\textbf{}}} & \multicolumn{6}{>{\columncolor{headerblue}}c|}{\textcolor{white}{\textbf{AudioCaps}}} & \multicolumn{6}{>{\columncolor{headerblue}}c|}{\textcolor{white}{\textbf{AVE-2}}} & \multicolumn{6}{>{\columncolor{headerblue}}c|}{\textcolor{white}{\textbf{TVSum}}} & \multicolumn{6}{>{\columncolor{headerblue}}c}{\textcolor{white}{\textbf{Ego4D}}} \\
\cmidrule(lr){2-7} \cmidrule(lr){8-13} \cmidrule(lr){14-19} \cmidrule(lr){20-25}
\multicolumn{1}{>{\columncolor{headerblue}}c|}{\textcolor{white}{}} & \multicolumn{3}{>{\columncolor{headerblue}}c|}{\textcolor{white}{\textbf{A→V}}} & \multicolumn{3}{>{\columncolor{headerblue}}c|}{\textcolor{white}{\textbf{V→A}}} & \multicolumn{3}{>{\columncolor{headerblue}}c|}{\textcolor{white}{\textbf{A→V}}} & \multicolumn{3}{>{\columncolor{headerblue}}c|}{\textcolor{white}{\textbf{V→A}}} & \multicolumn{3}{>{\columncolor{headerblue}}c|}{\textcolor{white}{\textbf{A→V}}} & \multicolumn{3}{>{\columncolor{headerblue}}c|}{\textcolor{white}{\textbf{V→A}}} & \multicolumn{3}{>{\columncolor{headerblue}}c|}{\textcolor{white}{\textbf{A→V}}} & \multicolumn{3}{>{\columncolor{headerblue}}c}{\textcolor{white}{\textbf{V→A}}} \\
\cmidrule(lr){2-4} \cmidrule(lr){5-7} \cmidrule(lr){8-10} \cmidrule(lr){11-13}
\cmidrule(lr){14-16} \cmidrule(lr){17-19} \cmidrule(lr){20-22} \cmidrule(lr){23-25}
\textcolor{white}{\textbf{Encoder}} & \textbf{T1} & \textbf{T3} & \textbf{T10} & \textbf{T1} & \textbf{T3} & \textbf{T10} & \textbf{T1} & \textbf{T3} & \textbf{T10} & \textbf{T1} & \textbf{T3} & \textbf{T10} & \textbf{T1} & \textbf{T3} & \textbf{T10} & \textbf{T1} & \textbf{T3} & \textbf{T10} & \textbf{T1} & \textbf{T3} & \textbf{T10} & \textbf{T1} & \textbf{T3} & \textbf{T10} \\
\midrule
\textcolor{darkgray}{\textbf{ImageBind}} & \textcolor{rawcolor}{\textbf{0.8}}→\textcolor{projectedcolor}{\textbf{45.2}} & \textcolor{rawcolor}{\textbf{4.0}}→\textcolor{projectedcolor}{\textbf{66.6}} & \textcolor{rawcolor}{\textbf{11.8}}→\textcolor{projectedcolor}{\textbf{84.0}} & \textcolor{rawcolor}{\textbf{1.4}}→\textcolor{projectedcolor}{\textbf{47.0}} & \textcolor{rawcolor}{\textbf{3.4}}→\textcolor{projectedcolor}{\textbf{68.0}} & \textcolor{rawcolor}{\textbf{11.4}}→\textcolor{projectedcolor}{\textbf{84.4}} & \textcolor{rawcolor}{\textbf{0.8}}→\textcolor{projectedcolor}{\textbf{40.0}} & \textcolor{rawcolor}{\textbf{2.4}}→\textcolor{projectedcolor}{\textbf{60.2}} & \textcolor{rawcolor}{\textbf{8.8}}→\textcolor{projectedcolor}{\textbf{76.2}} & \textcolor{rawcolor}{\textbf{0.2}}→\textcolor{projectedcolor}{\textbf{41.4}} & \textcolor{rawcolor}{\textbf{1.4}}→\textcolor{projectedcolor}{\textbf{60.0}} & \textcolor{rawcolor}{\textbf{9.2}}→\textcolor{projectedcolor}{\textbf{78.4}} & \textcolor{rawcolor}{\textbf{2.2}}→\textcolor{projectedcolor}{\textbf{19.2}} & \textcolor{rawcolor}{\textbf{4.6}}→\textcolor{projectedcolor}{\textbf{37.0}} & \textcolor{rawcolor}{\textbf{10.8}}→\textcolor{projectedcolor}{\textbf{62.8}} & \textcolor{rawcolor}{\textbf{1.0}}→\textcolor{projectedcolor}{\textbf{17.2}} & \textcolor{rawcolor}{\textbf{3.0}}→\textcolor{projectedcolor}{\textbf{38.2}} & \textcolor{rawcolor}{\textbf{9.6}}→\textcolor{projectedcolor}{\textbf{61.2}} & \textcolor{rawcolor}{\textbf{1.0}}→\textcolor{projectedcolor}{\textbf{8.6}} & \textcolor{rawcolor}{\textbf{2.6}}→\textcolor{projectedcolor}{\textbf{20.2}} & \textcolor{rawcolor}{\textbf{8.6}}→\textcolor{projectedcolor}{\textbf{40.4}} & \textcolor{rawcolor}{\textbf{0.6}}→\textcolor{projectedcolor}{\textbf{8.2}} & \textcolor{rawcolor}{\textbf{3.2}}→\textcolor{projectedcolor}{\textbf{14.8}} & \textcolor{rawcolor}{\textbf{8.6}}→\textcolor{projectedcolor}{\textbf{31.6}} \\
\rowcolor{lightimprovement}
\textcolor{darkgray}{\textbf{WhisperCLIP}} & \textcolor{rawcolor}{1.0}→\textcolor{projectedcolor}{18.6} & \textcolor{rawcolor}{3.2}→\textcolor{projectedcolor}{34.5} & \textcolor{rawcolor}{10.0}→\textcolor{projectedcolor}{62.3} & \textcolor{rawcolor}{1.4}→\textcolor{projectedcolor}{24.8} & \textcolor{rawcolor}{3.6}→\textcolor{projectedcolor}{41.2} & \textcolor{rawcolor}{10.4}→\textcolor{projectedcolor}{68.3} & \textcolor{rawcolor}{1.2}→\textcolor{projectedcolor}{21.8} & \textcolor{rawcolor}{3.2}→\textcolor{projectedcolor}{38.1} & \textcolor{rawcolor}{8.8}→\textcolor{projectedcolor}{64.2} & \textcolor{rawcolor}{1.0}→\textcolor{projectedcolor}{26.3} & \textcolor{rawcolor}{2.4}→\textcolor{projectedcolor}{43.6} & \textcolor{rawcolor}{10.8}→\textcolor{projectedcolor}{67.5} & \textcolor{rawcolor}{1.4}→\textcolor{projectedcolor}{13.6} & \textcolor{rawcolor}{2.8}→\textcolor{projectedcolor}{30.2} & \textcolor{rawcolor}{9.4}→\textcolor{projectedcolor}{59.0} & \textcolor{rawcolor}{1.8}→\textcolor{projectedcolor}{14.0} & \textcolor{rawcolor}{4.2}→\textcolor{projectedcolor}{31.8} & \textcolor{rawcolor}{10.6}→\textcolor{projectedcolor}{55.4} & \textcolor{rawcolor}{1.0}→\textcolor{projectedcolor}{7.0} & \textcolor{rawcolor}{3.2}→\textcolor{projectedcolor}{13.6} & \textcolor{rawcolor}{9.4}→\textcolor{projectedcolor}{28.4} & \textcolor{rawcolor}{1.2}→\textcolor{projectedcolor}{5.0} & \textcolor{rawcolor}{3.0}→\textcolor{projectedcolor}{12.8} & \textcolor{rawcolor}{9.0}→\textcolor{projectedcolor}{28.6} \\
\textcolor{darkgray}{\textbf{Wav2CLIP}} & \textcolor{rawcolor}{1.8}→\textcolor{projectedcolor}{13.0} & \textcolor{rawcolor}{4.6}→\textcolor{projectedcolor}{26.6} & \textcolor{rawcolor}{9.6}→\textcolor{projectedcolor}{44.6} & \textcolor{rawcolor}{1.6}→\textcolor{projectedcolor}{13.0} & \textcolor{rawcolor}{4.6}→\textcolor{projectedcolor}{26.4} & \textcolor{rawcolor}{11.6}→\textcolor{projectedcolor}{45.6} & \textcolor{rawcolor}{1.0}→\textcolor{projectedcolor}{14.6} & \textcolor{rawcolor}{3.4}→\textcolor{projectedcolor}{27.6} & \textcolor{rawcolor}{8.4}→\textcolor{projectedcolor}{47.8} & \textcolor{rawcolor}{0.8}→\textcolor{projectedcolor}{17.2} & \textcolor{rawcolor}{2.8}→\textcolor{projectedcolor}{28.4} & \textcolor{rawcolor}{9.8}→\textcolor{projectedcolor}{49.8} & \textcolor{rawcolor}{1.4}→\textcolor{projectedcolor}{5.6} & \textcolor{rawcolor}{2.4}→\textcolor{projectedcolor}{15.2} & \textcolor{rawcolor}{11.8}→\textcolor{projectedcolor}{36.4} & \textcolor{rawcolor}{0.4}→\textcolor{projectedcolor}{5.0} & \textcolor{rawcolor}{2.8}→\textcolor{projectedcolor}{16.4} & \textcolor{rawcolor}{11.4}→\textcolor{projectedcolor}{37.8} & \textcolor{rawcolor}{0.8}→\textcolor{projectedcolor}{3.6} & \textcolor{rawcolor}{1.8}→\textcolor{projectedcolor}{7.4} & \textcolor{rawcolor}{8.0}→\textcolor{projectedcolor}{17.6} & \textcolor{rawcolor}{0.6}→\textcolor{projectedcolor}{1.8} & \textcolor{rawcolor}{2.6}→\textcolor{projectedcolor}{6.0} & \textcolor{rawcolor}{9.2}→\textcolor{projectedcolor}{16.4} \\
\rowcolor{lightimprovement}
\textcolor{darkgray}{\textbf{AudioCLIP}} & \textcolor{rawcolor}{1.0}→\textcolor{projectedcolor}{6.4} & \textcolor{rawcolor}{3.6}→\textcolor{projectedcolor}{14.4} & \textcolor{rawcolor}{9.8}→\textcolor{projectedcolor}{35.4} & \textcolor{rawcolor}{1.6}→\textcolor{projectedcolor}{7.4} & \textcolor{rawcolor}{4.8}→\textcolor{projectedcolor}{18.2} & \textcolor{rawcolor}{11.0}→\textcolor{projectedcolor}{37.8} & \textcolor{rawcolor}{0.8}→\textcolor{projectedcolor}{8.2} & \textcolor{rawcolor}{2.8}→\textcolor{projectedcolor}{17.8} & \textcolor{rawcolor}{9.2}→\textcolor{projectedcolor}{38.8} & \textcolor{rawcolor}{1.0}→\textcolor{projectedcolor}{11.0} & \textcolor{rawcolor}{3.8}→\textcolor{projectedcolor}{23.0} & \textcolor{rawcolor}{13.6}→\textcolor{projectedcolor}{45.0} & \textcolor{rawcolor}{1.0}→\textcolor{projectedcolor}{7.2} & \textcolor{rawcolor}{2.8}→\textcolor{projectedcolor}{17.2} & \textcolor{rawcolor}{10.2}→\textcolor{projectedcolor}{35.6} & \textcolor{rawcolor}{1.3}→\textcolor{projectedcolor}{6.8} & \textcolor{rawcolor}{2.3}→\textcolor{projectedcolor}{14.4} & \textcolor{rawcolor}{8.8}→\textcolor{projectedcolor}{33.6} & \textcolor{rawcolor}{1.4}→\textcolor{projectedcolor}{2.4} & \textcolor{rawcolor}{3.0}→\textcolor{projectedcolor}{6.8} & \textcolor{rawcolor}{10.0}→\textcolor{projectedcolor}{15.6} & \textcolor{rawcolor}{1.4}→\textcolor{projectedcolor}{1.0} & \textcolor{rawcolor}{3.2}→\textcolor{projectedcolor}{4.8} & \textcolor{rawcolor}{10.6}→\textcolor{projectedcolor}{14.0} \\
\textcolor{darkgray}{\textbf{CLAP}} & \textcolor{rawcolor}{1.4}→\textcolor{projectedcolor}{6.8} & \textcolor{rawcolor}{2.8}→\textcolor{projectedcolor}{18.4} & \textcolor{rawcolor}{10.2}→\textcolor{projectedcolor}{38.6} & \textcolor{rawcolor}{1.4}→\textcolor{projectedcolor}{10.4} & \textcolor{rawcolor}{2.6}→\textcolor{projectedcolor}{20.0} & \textcolor{rawcolor}{8.6}→\textcolor{projectedcolor}{41.0} & \textcolor{rawcolor}{0.6}→\textcolor{projectedcolor}{7.2} & \textcolor{rawcolor}{2.2}→\textcolor{projectedcolor}{16.2} & \textcolor{rawcolor}{9.6}→\textcolor{projectedcolor}{36.8} & \textcolor{rawcolor}{0.8}→\textcolor{projectedcolor}{10.2} & \textcolor{rawcolor}{3.4}→\textcolor{projectedcolor}{23.2} & \textcolor{rawcolor}{10.8}→\textcolor{projectedcolor}{43.4} & \textcolor{rawcolor}{0.6}→\textcolor{projectedcolor}{2.6} & \textcolor{rawcolor}{2.4}→\textcolor{projectedcolor}{7.2} & \textcolor{rawcolor}{10.2}→\textcolor{projectedcolor}{22.6} & \textcolor{rawcolor}{1.2}→\textcolor{projectedcolor}{3.4} & \textcolor{rawcolor}{2.8}→\textcolor{projectedcolor}{10.4} & \textcolor{rawcolor}{8.2}→\textcolor{projectedcolor}{25.0} & \textcolor{rawcolor}{0.6}→\textcolor{projectedcolor}{2.6} & \textcolor{rawcolor}{1.6}→\textcolor{projectedcolor}{8.8} & \textcolor{rawcolor}{9.4}→\textcolor{projectedcolor}{22.6} & \textcolor{rawcolor}{1.2}→\textcolor{projectedcolor}{3.2} & \textcolor{rawcolor}{4.2}→\textcolor{projectedcolor}{6.4} & \textcolor{rawcolor}{14.0}→\textcolor{projectedcolor}{19.8} \\
\bottomrule
\end{tabular}
}
\caption{\textbf{Cross-modal retrieval results with raw→projected comparison (\%).} Shows performance improvement from raw embeddings to projected across all datasets and metrics. T1/T3/T10 denote Top-1/3/10 accuracy for Audio→Video (A→V) and Video→Audio (V→A) retrieval. \textcolor{rawcolor}{\textbf{Red values}}: raw embeddings, \textcolor{projectedcolor}{\textbf{green values}}: projected embeddings. }
\label{tab:retrieval_complete_comparison}
\end{table*}

To systematically investigate the retrieval–generation trade-off in audio–visual representation learning, we conducted extensive experiments that included multiple audio encoders, datasets, and evaluation protocols. Our experimental design specifically targets two key questions: (1) How effectively can different audio encoders be aligned with CLIP's visual space? and (2) What are the implications of such alignment for downstream tasks?

\subsection{Experimental Setup}
\label{subsec:setup}

We evaluate on four diverse datasets to ensure robust findings across different domains: (1) AVE-2, our primary benchmark containing 580{,}145 audio–visual clips with fine-grained alignment annotations and is a subset of AudioSet \citep{audioset}; (2) AudioCaps \citep{kim2019audiocaps}, featuring audio captions for AudioSet segments; (3) TVSum \citep{song2015tvsum}, containing summarization-focused video content; and (4) Ego4D \citep{grauman2022ego4d}, representing challenging egocentric perspectives.

We comprehensively evaluate five state-of-the-art audio representation models: (1) Whisper \citep{radford2023whisper}, an ASR-focused encoder trained on 680K hours of supervised data; (2) ImageBind \citep{girdhar2023imagebind}, a joint embedding model trained on image-text-audio triplets; (3) CLAP \citep{elizalde2023clap}, trained on audio-text pairs; (4) AudioCLIP \citep{guzhov2022audioclip}, which extends CLIP's contrastive learning to audio; and (5) Wav2CLIP \citep{wu2022wav2clip}, which distills knowledge from CLIP's visual space to audio. Additionally, we evaluate our proposed WhisperCLIP, which enhances Whisper through multi-layer fusion and projecting it to the representation space of the CLIP.

We assess performance across two key areas: \textit{cross-modal retrieval} aligned with CLIP and training-free text generation with visual token substitution. For retrieval, we measure Top-$\{1,3,10\}$ accuracy for both Audio→Video and Video→Audio retrievals. For generation quality, we use both conventional NLP metrics, such as BLEU, ROUGE-L, BERTScore, CLIP-Score, and LLM as judge using GPT-4o along six dimensions, including audio quality, visual accuracy, coherence, reasoning, fluency, and overall usefulness.

The SoundCLIP projection networks were structured as three-layer MLPs with LayerNorm and GELU activations (1.9M parameters). For token substitution, we investigate two budget options: $k=15$ (audio-driven) and $k=150$ (vision-driven). The numbers 15 and 150 were based on empirical study to obtain the minimum critical number of tokens when audio tokens or visual tokens become dominant.

\subsection{Cross-Modal Retrieval Experiments}
\label{subsec:retrieval_experiments}

For each encoder and dataset, we evaluate the top accuracies and mean retrieval ranks, both with and without alignment to CLIP. Table~\ref{tab:retrieval_complete_comparison} illustrates the retrieval performance using raw embeddings as well as when projections are aligned to CLIP's visual space.
Across all datasets, projecting the audio features into CLIP's space yields large improvements in retrieval accuracy. The effect is especially dramatic for image-centric encoders: ImageBind's Top-1 audio-to-video retrieval on AudioCaps improves from 0.8\% to 45.2\%, while Wav2CLIP and AudioCLIP see more modest but still substantial gains. WhisperCLIP, despite already being trained on language, benefits from projection almost as much as vision-centric encoders, indicating that even language-supervised models contain information that CLIP alignment can exploit. We also observe that vision-aligned models generally achieve higher raw retrieval scores on vision-heavy datasets like TVSum and Ego4D, while language-supervised models do better on caption-rich AudioCaps.

\subsection{Generation Quality Experiments}
\label{subsec:generation_experiments}

We evaluate text generation quality with raw versus CLIP-aligned embeddings. Table~\ref{tab:generation_quality} validates our claim that CLIP-aligned embeddings degrade generation quality across all metrics for $k=150$.  WhisperCLIP produces the highest scores across BLEU, ROUGE and BERTScore when using raw embeddings, with CLAP and ImageBind close behind, highlighting the benefit of language supervision.  AudioCLIP and Wav2CLIP trail slightly, reflecting the fact that their training on vision features does not embed as much linguistic information.  The drop in performance when aligning to CLIP is consistent across encoders: BLEU scores fall by roughly 20--25,\% on average, and GPT‑4o evaluations decline by 0.3–0.4 points.  This pattern corroborates our assertion that tighter alignment sacrifices semantic richness needed for fluent captions.

\begin{table}[htbp]
\centering
\caption{\textbf{Generation quality evaluation.} Raw embeddings surpass CLIP-aligned variants in all encoders.}
\label{tab:generation_quality}
\scalebox{0.85}{
\footnotesize
\setlength{\tabcolsep}{4pt}
\renewcommand{\arraystretch}{1.1}
\begin{tabular}{l>{\columncolor{blue!25}}c|cccc|>{\columncolor{gray!25}}c}
\toprule
\scriptsize\textbf{Encoder} & \scriptsize\textbf{CLIP} & \scriptsize\textbf{BLEU} & \scriptsize\textbf{ROUGE} & \scriptsize\textbf{BERT} & \scriptsize\textbf{CLIP} & \scriptsize\textbf{GPT-4o} \\
 & \scriptsize\textbf{aligned} & \scriptsize\textbf{No/Yes} & \scriptsize\textbf{No/Yes} & \scriptsize\textbf{No/Yes} & \scriptsize\textbf{No/Yes} & \scriptsize\textbf{No/Yes} \\
\midrule
\scriptsize WhisperCLIP & \scriptsize No/Yes & \scriptsize\textbf{0.103}/0.078 & \scriptsize\textbf{0.284}/0.263 & \scriptsize\textbf{0.271}/0.219 & \scriptsize\textbf{0.782}/0.715 & \scriptsize\textbf{1.93}/1.58 \\
\scriptsize ImageBind & \scriptsize No/Yes & \scriptsize\textbf{0.096}/0.076 & \scriptsize\textbf{0.278}/0.254 & \scriptsize\textbf{0.267}/0.210 & \scriptsize\textbf{0.775}/0.719 & \scriptsize\textbf{1.89}/1.53 \\
\scriptsize CLAP & \scriptsize No/Yes & \scriptsize\textbf{0.096}/0.078 & \scriptsize\textbf{0.278}/0.256 & \scriptsize\textbf{0.268}/0.220 & \scriptsize\textbf{0.776}/0.717 & \scriptsize\textbf{1.92}/1.55 \\
\scriptsize AudioCLIP & \scriptsize No/Yes & \scriptsize\textbf{0.093}/0.080 & \scriptsize\textbf{0.274}/0.260 & \scriptsize\textbf{0.260}/0.224 & \scriptsize\textbf{0.768}/0.727 & \scriptsize\textbf{1.81}/1.64 \\
\scriptsize Wav2CLIP & \scriptsize No/Yes & \scriptsize\textbf{0.095}/0.085 & \scriptsize\textbf{0.268}/0.265 & \scriptsize\textbf{0.247}/0.238 & \scriptsize\textbf{0.754}/0.742 & \scriptsize\textbf{1.80}/1.67 \\
\bottomrule
\end{tabular}
}
\end{table}

Raw embeddings consistently produce higher-quality text generation compared to their CLIP-aligned counterparts across all encoders and metrics. WhisperCLIP achieves the best generation performance in both raw (BLEU 0.103, BERTScore 0.271, GPT-4o 1.93) and CLIP-aligned form (BLEU 0.078, BERTScore 0.219, GPT-4o 1.58).

Table~\ref{tab:generation_quality_comparison} breaks down the impact of alignment across five dimensions. Alignment most severely diminishes audio fidelity and reasoning, with average degradations of 0.16 and 0.34 points, respectively.  The smallest drop occurs in fluency ($\approx 0.12$), suggesting that syntactic quality is largely preserved even when semantic richness is reduced.  WhisperCLIP and CLAP retain the highest scores across dimensions, while vision-centric encoders suffer larger degradations, underscoring the advantage of language supervision for open-ended generation.

\begin{table}[htbp]
\centering
\caption{\textbf{Generation quality comparison: Raw vs CLIP-aligned embeddings across evaluation dimensions.} Format: Raw/Aligned (+Gain). }
\label{tab:generation_quality_comparison}
\scalebox{0.56}{
\footnotesize
\setlength{\tabcolsep}{4pt}
\renewcommand{\arraystretch}{1.1}
\begin{tabular}{l|>{\columncolor{blue!25}}c>{\columncolor{blue!25}}c>{\columncolor{blue!25}}c>{\columncolor{blue!25}}c|>{\columncolor{gray!25}}c}
\toprule
\textbf{Encoder} & \textbf{Audio} & \textbf{Visual} & \textbf{Coherence} & \textbf{Reasoning} & \textbf{Fluency} \\
 & \textbf{Raw/Aligned} & \textbf{Raw/Aligned} & \textbf{Raw/Aligned} & \textbf{Raw/Aligned} & \textbf{Raw/Aligned} \\
\midrule
CLAP & \textbf{1.52}/1.29 (+0.23) & \textbf{2.59}/1.98 (+0.61) & \textbf{1.63}/1.36 (+0.27) & \textbf{2.21}/1.76 (+0.45) & \textbf{4.05}/3.89 (+0.16) \\
WhisperCLIP & \textbf{1.49}/1.31 (+0.18) & \textbf{2.49}/1.84 (+0.65) & \textbf{1.60}/1.32 (+0.28) & \textbf{2.13}/1.69 (+0.44) & \textbf{4.00}/3.87 (+0.13) \\
Wav2CLIP & \textbf{1.48}/1.34 (+0.14) & \textbf{2.33}/2.17 (+0.16) & \textbf{1.57}/1.43 (+0.14) & \textbf{2.05}/1.90 (+0.15) & \textbf{3.99}/3.95 (+0.04) \\
AudioCLIP & \textbf{1.44}/1.39 (+0.05) & \textbf{2.39}/2.05 (+0.34) & \textbf{1.56}/1.43 (+0.13) & \textbf{2.06}/1.86 (+0.20) & \textbf{3.99}/3.90 (+0.09) \\
ImageBind & \textbf{1.50}/1.29 (+0.21) & \textbf{2.52}/1.92 (+0.60) & \textbf{1.64}/1.35 (+0.29) & \textbf{2.16}/1.72 (+0.44) & \textbf{4.04}/3.86 (+0.18) \\
\midrule
\textbf{Avg. Gain} & \textbf{+0.16} & \textbf{+0.47} & \textbf{+0.22} & \textbf{+0.34} & \textbf{+0.12} \\
\bottomrule
\end{tabular}
}
\end{table}

\begin{figure}[h]
    \centering
    \scalebox{0.47}{
    \begin{tabular}{c@{\hspace{2mm}} c@{\hspace{1mm}} c@{\hspace{1mm}} c@{\hspace{1mm}} c@{\hspace{1mm}} c}
        & \textbf{ImageBind} &
        \textbf{AudioCLIP} &
        \textbf{CLAP} &
        \textbf{Wav2CLIP} &
        \textbf{WhisperCLIP} \\[2pt]
        
        \rotatebox{90}{\textbf{Raw}} &
        \includegraphics[width=0.18\textwidth]{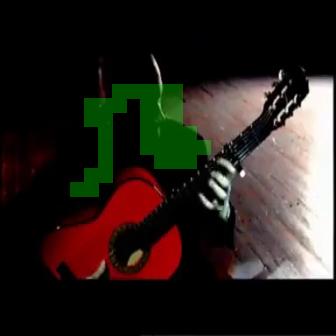} &
        \includegraphics[width=0.18\textwidth]{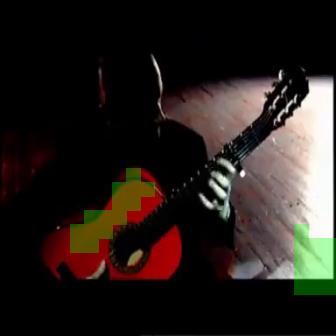} &
        \includegraphics[width=0.18\textwidth]{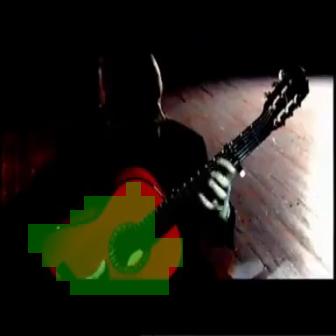} &
        \includegraphics[width=0.18\textwidth]{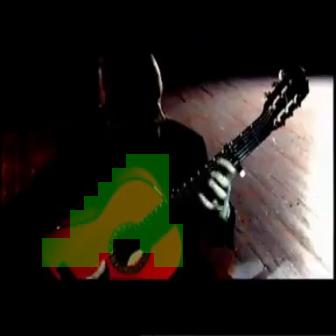} &
        \includegraphics[width=0.18\textwidth]{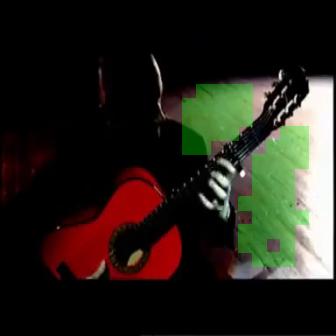} \\[1pt]
        
        \rotatebox{90}{\textbf{Projected}} &
        \includegraphics[width=0.18\textwidth]{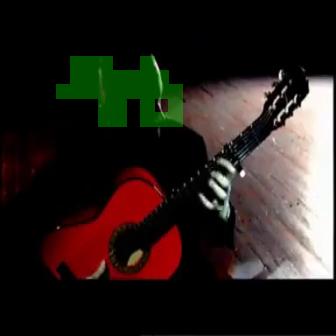} &
        \includegraphics[width=0.18\textwidth]{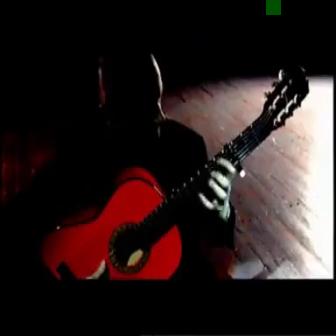} &
        \includegraphics[width=0.18\textwidth]{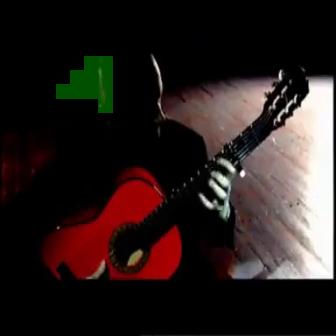} &
        \includegraphics[width=0.18\textwidth]{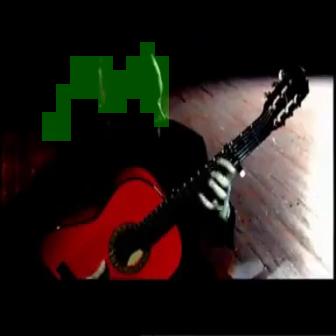} &
        \includegraphics[width=0.18\textwidth]{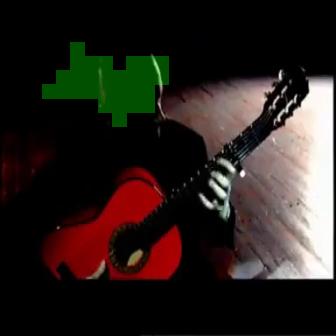} \\[5pt]
        
        \rotatebox{90}{\textbf{Raw}} &
        \includegraphics[width=0.18\textwidth]{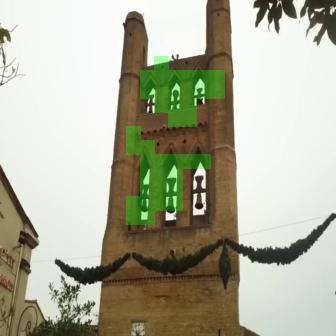} &
        \includegraphics[width=0.18\textwidth]{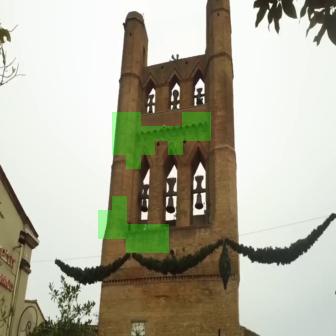} &
        \includegraphics[width=0.18\textwidth]{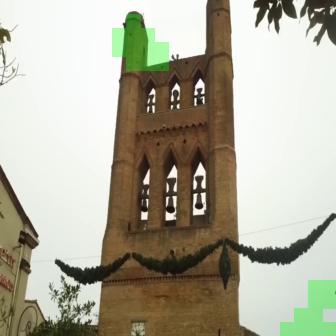} &
        \includegraphics[width=0.18\textwidth]{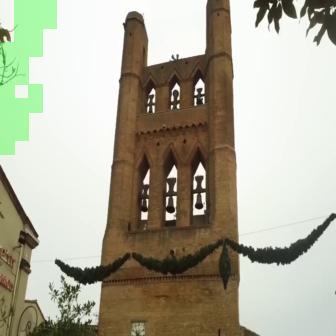} &
        \includegraphics[width=0.18\textwidth]{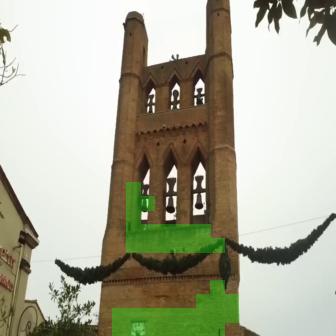} \\[1pt]
        
        \rotatebox{90}{\textbf{Projected}} &
        \includegraphics[width=0.18\textwidth]{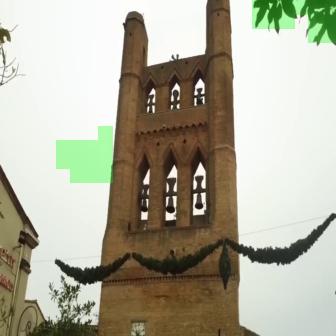} &
        \includegraphics[width=0.18\textwidth]{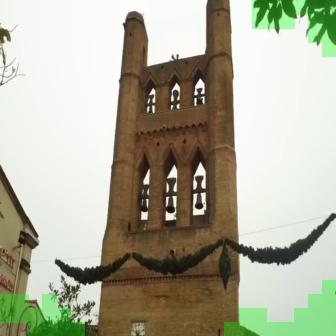} &
        \includegraphics[width=0.18\textwidth]{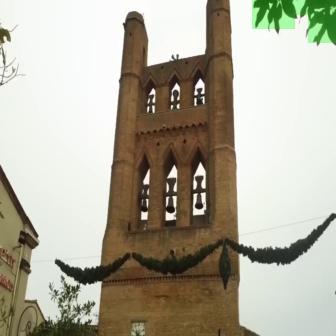} &
        \includegraphics[width=0.18\textwidth]{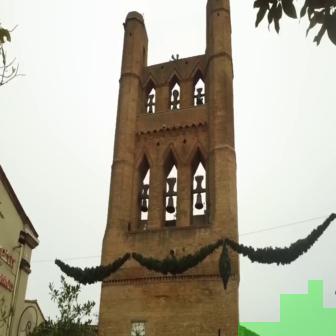} &
        \includegraphics[width=0.18\textwidth]{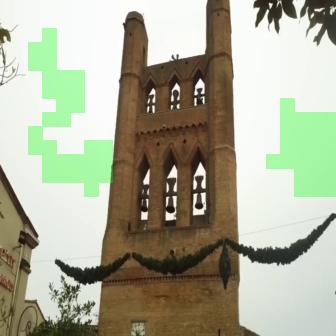} \\[2pt]
    \end{tabular}
    }
    \caption{Comparison of audio-guided patch selection using \textbf{raw} vs. \textbf{projected} embeddings. Bright green regions indicate the 100 most-attended ViT patches selected by each encoder. Top rows show a guitar performance scene, bottom rows show a church bell.}
    \label{fig:qual_results_combined}
\end{figure}

We visualized the most attended visual patches for different encoders in both raw and projected configurations, as shown in Fig. \ref{fig:qual_results_combined}.  Raw embeddings produce more distributed attention patterns that capture both sound sources and their context, while projected embeddings focus more narrowly on sound-producing objects.  These attention maps further illustrate why raw embeddings support better generation: they attend to both the sound source and its contextual surroundings, while aligned embeddings concentrate almost exclusively on the sound-producing object, missing broader contextual cues.  This visualization helps explain why raw embeddings yield more contextually rich descriptions despite lower retrieval.

\subsection{Token Budget Analysis}
\label{subsec:token_budget}

We repeat generation experiments with small ($k=15$) and large ($k=150$) token budgets for each encoder. Table~\ref{tab:token_budget_analysis} illustrates that more visual tokens improve generation quality but do not eliminate the gap between raw and aligned embeddings.  Increasing the number of visual tokens from 15 to 150 benefits all models, with GPT‑4o scores rising by roughly 0.4 points on average.  However, raw embeddings remain superior to aligned ones even with larger token budgets, reinforcing that the loss of semantic richness cannot be compensated solely by adding more visual context.  Notably, CLAP and ImageBind both achieve GPT‑4o scores near 1.9 when $k=150$ in raw mode, whereas their aligned counterparts lag behind by 0.3–0.4 points.

\begin{table}[htbp]
\centering
\caption{Impact of visual token budget ($k$) on generation quality across encoders (GPT-4o overall scores).}
\label{tab:token_budget_analysis}
\scalebox{0.7}{
\renewcommand{\arraystretch}{1.2}
\begin{tabular}{l|>{\columncolor{blue!25}}c>{\columncolor{blue!25}}c|>{\columncolor{gray!25}}c>{\columncolor{gray!25}}c}
\toprule
\multirow{2}{*}{\textbf{Encoder}} & \multicolumn{2}{c|}{\textbf{$k$=15}} & \multicolumn{2}{c}{\textbf{$k$=150}} \\
& Raw & Aligned & Raw & Aligned \\
\midrule
CLAP & 1.47±0.05 & 1.15±0.03 & \textbf{1.92±0.06} & 1.55±0.05 \\
WhisperCLIP & 1.41±0.05 & 1.14±0.03 & \textbf{1.87±0.06} & 1.50±0.05 \\
Wav2CLIP & 1.34±0.04 & 1.21±0.03 & \textbf{1.80±0.06} & 1.67±0.05 \\
AudioCLIP & 1.31±0.04 & 1.18±0.03 & \textbf{1.81±0.05} & 1.64±0.05 \\
ImageBind & 1.40±0.05 & 1.17±0.03 & \textbf{1.89±0.06} & 1.53±0.05 \\
\bottomrule
\end{tabular}
}
\end{table}

We group encoders into “language-supervised” (Whisper, CLAP) and “vision-aligned” (AudioCLIP, Wav2CLIP, ImageBind) categories. Note that ImageBind is trained on multiple modalities including textual descriptions but uses image representations as the central modality, hence we categorize it as vision-aligned. Table~\ref{tab:training_paradigm_comparison} presents their GPT‑4o scores side by side.  Encoders with language exposure during pretraining show competitive generation capabilities in their raw form, with language-supervised encoders achieving slightly higher average scores (1.90) compared to vision-aligned encoders (1.83), confirming that exposure to text during training confers advantages for generation.

\begin{table}[htbp]
\centering
\caption{Generation performance by training paradigm (raw embeddings, $k=150$). BLEU scores range 0-1, BERTScore range 0-1, GPT-4o scores range 1-5 (higher better).}
\label{tab:training_paradigm_comparison}
\scalebox{0.6}{
\renewcommand{\arraystretch}{1.2}
\begin{tabular}{l|>{\columncolor{blue!25}}l|cc>{\columncolor{gray!25}}c}
\toprule
\textbf{Training Type} & \textbf{Encoder} & \textbf{GPT-4o} & \textbf{BLEU} & \textbf{BERTScore} \\
\midrule
\multirow{2}{*}{\cellcolor{white}\textbf{Language-Supervised}} & CLAP & \textbf{1.92} & 0.096 & 0.268 \\
& Whisper & 1.87 & \textbf{0.097} & 0.265 \\
\midrule
\multirow{3}{*}{\cellcolor{white}\textbf{Vision-Aligned}} & AudioCLIP & 1.81 & 0.093 & 0.260 \\
& Wav2CLIP & 1.80 & 0.095 & 0.247 \\
& ImageBind & 1.89 & 0.096 & \textbf{0.267} \\
\bottomrule
\end{tabular}
}
\end{table}
\subsection{WhisperCLIP Ablations}
\label{subsec:whisperclip_ablations}
For loss function comparison, we train WhisperCLIP’s projection MLP using different losses. Table~\ref{tab:loss_function_ablation} tabulates mean retrieval ranks to identify the best training objective.  Among the losses tested, combining InfoNCE with distribution matching yields the lowest mean retrieval rank, outperforming simple cosine or MSE losses by a wide margin.  This suggests that both contrastive alignment and distribution regularization are important for learning effective audio-to-visual projections.

\begin{table}[htbp]
\centering
\caption{Impact of different loss functions on retrieval performance (mean rank, lower is better).}
\label{tab:loss_function_ablation}
\scalebox{0.6}{
\renewcommand{\arraystretch}{1.2}
\begin{tabular}{>{\columncolor{blue!25}}l|cc}
\toprule
\textbf{Loss Function} & \textbf{A→V Mean Rank} & \textbf{V→A Mean Rank} \\
\midrule
Cosine Similarity & 39.1 & 18.8 \\
MSE Loss & 48.3 & 38.5 \\
Distribution Matching & 37.8 & 16.6 \\
InfoNCE & 14.3 & 12.2 \\
\textbf{InfoNCE + Distribution Match} & \textbf{12.4} & \textbf{11.6} \\
\bottomrule
\end{tabular}
}
\end{table}

Table~\ref{tab:pooling_strategy_ablation} evaluates retrieval performance when using different pooling strategies.  Averaging across all Whisper layers provides the strongest performance, surpassing final-layer or weighted-sum approaches.  This supports our hypothesis that retaining information from multiple layers captures both lower-level acoustic details and higher-level semantic content, and thus we adopt layer averaging in our main experiments.

\begin{table}[!t]
\centering
\caption{Impact of various pooling strategies on performance.}
\label{tab:pooling_strategy_ablation}
\scalebox{0.6}{
\renewcommand{\arraystretch}{1.2}
\begin{tabular}{>{\columncolor{blue!25}}l|c>{\columncolor{gray!25}}c}
\toprule
\textbf{Pooling Strategy} & \textbf{A→V Mean Rank} & \textbf{V→A Mean Rank} \\
\midrule
Final-layer only & 38.8 & 19.7 \\
Middle-layer only & 28.3 & 23.2 \\
Last-$n$ layers ($n=2$) & 27.4 & 16.4 \\
Last-$n$ layers ($n=3$) & 27.4 & 16.4 \\
Weighted-sum-all layers & 13.6 & 12.0 \\
\textbf{Average-all layers} & \textbf{11.4} & \textbf{10.7} \\
\bottomrule
\end{tabular}
}
\end{table}

Table~\ref{tab:backbone_size_ablation} compares retrieval ranks in different Whisper backbone sizes.  Larger backbones consistently yield better retrieval ranks, with the medium configuration providing the best trade-off between performance and parameter count.  We therefore use the medium-size Whisper encoder as our default choice.  Table~\ref{tab:last_n_layer_ablation} further shows that using only the last three Whisper layers [of tiny Whisper] outperforms models that use only the last one or two layers.

\begin{table}[htbp]
\centering
\caption{Impact of Whisper backbone size on performance.}
\label{tab:backbone_size_ablation}
\scalebox{0.6}{
\renewcommand{\arraystretch}{1.2}
\begin{tabular}{>{\columncolor{blue!25}}l|cc|>{\columncolor{gray!25}}c}
\toprule
\textbf{Whisper Size} & \textbf{A→V Mean Rank} & \textbf{V→A Mean Rank} & \textbf{Parameters} \\
\midrule
Tiny & 26.4 & 17.5 & 39M \\
Base & 12.2 & 11.5 & 74M \\
Small & 10.8 & 9.4 & 244M \\
\textbf{Medium} & \textbf{10.3} & \textbf{8.7} & 769M \\
\bottomrule
\end{tabular}
}
\end{table}


\begin{table}[htbp]
\centering
\caption{Last-$n$-layers ablation (InfoNCE loss).}
\label{tab:last_n_layer_ablation}
\scalebox{0.75}{
\renewcommand{\arraystretch}{1.2}
\begin{tabular}{>{\columncolor{blue!25}}l|c>{\columncolor{gray!25}}c}
\toprule
\textbf{Layers Used} & \textbf{A→V Mean Rank} & \textbf{V→A Mean Rank} \\
\midrule
Last 1 & 38.8 & 19.7 \\
Last 2 & 37.1 & 23.8 \\
\textbf{Last 3} & \textbf{27.4} & \textbf{16.4} \\
\bottomrule
\end{tabular}
}
\end{table}
\subsection{Retrieval–Generation Trade-off Analysis}
\label{subsec:tradeoff_analysis}

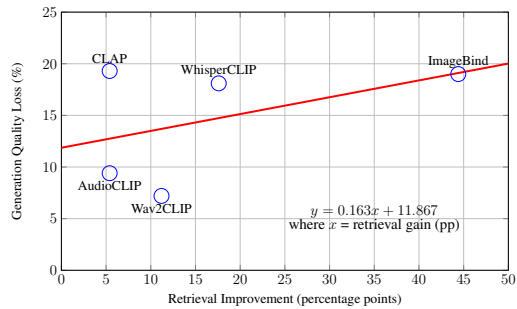
\begin{figure}[htbp]
    \centering
    \scalebox{0.47}{
    \begin{tikzpicture}
        \begin{axis}[
            width=0.8\textwidth,
            height=0.5\textwidth,
            xlabel={Retrieval Improvement (percentage points)},
            ylabel={Generation Quality Loss (\%)},
            xlabel style={font=\large},
            ylabel style={font=\large},
            tick label style={font=\Large},
            grid=both,
            xmin=0, xmax=50,
            ymin=0, ymax=25,
            ]
            \addplot[mark=o, blue, mark size=6pt, only marks] coordinates {
                (5.4, 19.3) 
                (5.4, 9.4)  
                (11.2, 7.2) 
                (17.6, 18.1) 
                (44.4, 19.0) 
            };
            \node at (axis cs:5.4,20.5) {\large CLAP};
            \node at (axis cs:5.4,8.2) {\large AudioCLIP};
            \node at (axis cs:11.2,6) {\large Wav2CLIP};
            \node at (axis cs:17.6,19.3) {\large WhisperCLIP};
            \node at (axis cs:44.4,20.2) {\large ImageBind};
            \addplot[red, ultra thick, domain=0:50] {0.163*x + 11.867};
            \node at (axis cs:35,5) [align=center] {\Large $y = 0.163x + 11.867$ \\ \Large where $x$ = retrieval gain (pp)};
        \end{axis}
    \end{tikzpicture}
    }
    \caption{\textbf{The fundamental retrieval-generation trade-off.} Positive correlation ($r \approx 0.45$) between retrieval improvement and generation quality loss across all encoders.}
    \label{fig:tradeoff_scatter}
\end{figure}

For each encoder, we compute the percentage-point increase in Top-1 retrieval accuracy due to alignment and the corresponding decrease in GPT‑4o generation score.  Figure~\ref{fig:tradeoff_scatter} plots each encoder, revealing a positive correlation of $r \approx 0.45$.  We perform least-squares linear regression to quantify this relationship, yielding the trend line
[
y = 0.163,x + 11.867,
]
where $x$ represents the improvement in retrieval percentage points and $y$ represents the loss of generation quality as a percentage.  This equation indicates that each percentage-point gain in retrieval performance incurs approximately 0.163\% loss in generation quality, with a baseline degradation of 11.867\% even at minimal retrieval improvements.  Wav2CLIP shows the most favorable trade-off ratio (0.64\% generation loss per retrieval point), while CLAP exhibits the highest cost (3.57\% loss per point), illustrating how model architecture and supervision paradigm influence the Pareto frontier.

\section{Related Work}
\label{sec:related_works}

Our research extends previous work on multisensory perception and multimodal representation learning. Early studies, such as the McGurk effect, demonstrated how visual cues can alter auditory perceptions~\citep{tiippana2014mcgurk}. Further research linked auditory and visual elements, aiding in the development of models such as Hidden Markov Models for better audio-visual synchronization~\citep{evans2010natural, bengio2002ahmm}. The domains of multimodal understanding and generation have garnered considerable scholarly interest over the past decade in a range of fields, including augmented reality \citep{bi2025i2g, bi2023misar, bi2024eagle,tang2025mmperspective, nguyen2024oscar}, personalization \citep{tang2025catv}, scientific discovery and verification \citep{vosoughi2025openxrd, bi2025verify}, as well as issues related to bias and reliability \citep{mozaffari2025glen,vosoughi2024cross}, language, and audio, among others \citep{vosoughi2024counterfactual, su2023avsa}. These efforts have significantly improved cross-modal perception. Recent advancements in multimodal pretraining have enhanced multi-domain alignment. Studies on text-free audio-visual learning emphasize maintaining natural acoustic details~\citep{radford2021learning, liu2025intentionalgesturedeliverintentions,elizalde2023clap, girdhar2023imagebind, hamilton2024separating, liu2025contextualgesturecospeechgesture, gtr, arandjelovic2023objectsounds, liu2023visual, bi2025reasoningmatterssurveyadvancements, li2023blip, ashutosh2025llms}. Foundation models such as AudioCLIP, Wav2CLIP, ImageBind, CLAP, and Whisper have linked audio to visual and textual data~\citep{guzhov2022audioclip, wu2022wav2clip, girdhar2023imagebind, elizalde2023clap, radford2023whisper}. EchoInk‑R1, MLAVL, CCNet, LoCo, AVHBench, AVSGN, UWAV, AvED, and AVEdit are some of the key developments focusing on improving cross-modal interactions and applications in multimedia~\citep{radford2021learning, guzhov2022audioclip, wu2022wav2clip, girdhar2023imagebind, elizalde2023clap, radford2023whisper, ma2024maavt, duan2023dgsct, xing2025echoink, xu2025mlavl, zhou2025ccnet, xing2025loco, kim2025avhbench, liu2025avsgn, lai2025uwav, lin2025zero, liang2025avedit}.


\section{Conclusion}
\label{sec:conclusion}

We have investigated the effects of aligning audio embeddings with a fixed visual manifold, utilizing the new AVE‑2 dataset and the innovative \frameworkname{} integration approach. Our comprehensive analysis included five audio encoders from both language-supervised and vision-centric frameworks. The results indicate a significant trade-off: while alignment with the CLIP visual space enhances retrieval capabilities, it compromises the quality of generation. Language-supervised encoders, such as Whisper and CLAP, preserve more semantic content compared to vision-centric encoders. Our fusion strategy, WhisperCLIP, achieves an optimal balance between retrieval efficiency and generation quality. These insights delineate a Pareto frontier in cross-modal representation learning and offer specific recommendations for selecting audio encoders based on project needs. Future research could broaden this study to additional modalities and investigate training methodologies that simultaneously improve retrieval and generative accuracy.

\section*{Acknowledgment}
The authors express gratitude to Dimitra Emmanouilidou and Hannes Gamper from Microsoft Research for their valuable discussions that contributed to the enrichment of this work.

\bibliography{references}

\end{document}